\documentclass[aps,pra,
  twocolumn,
  longbibliography
]{revtex4-2}
\usepackage{amsmath,amssymb,amsfonts,amsthm}
\usepackage{bm,mathrsfs,mathtools}
\usepackage[normalem]{ulem}
\usepackage{dsfont}
\usepackage{textcomp}
\usepackage{upgreek,textgreek}
\usepackage[
  colorlinks=true,
  urlcolor=magenta,
  linkcolor=blue,
  anchorcolor=blue,
  citecolor=blue
]{hyperref}
\usepackage{physics}
\usepackage{orcidlink}
\usepackage{graphicx}
\usepackage{float}
\usepackage{siunitx}
\usepackage{booktabs}
\usepackage{color}
\usepackage{graphicx}
\usepackage[caption=false]{subfig}
\begin{document}

\title{\textbf{The influence of the transverse electric field on accelerating vortex state in the axisymmetric electric field}}

\author{Ziyang Ding$^{1}$}
\author{Ziqiang Huang$^{2}$}
\author{Qi Meng$^{2}$}
\author{Alexander J. Silenko$^{3,4}$}
\author{Pengming Zhang$^{1}$}
\author{Liping Zou$^{2}$}
%\email[Contact author:~]{zoulp5@mail.sysu.edu.cn}
%
\affiliation{$^1$School of Physics and Astronomy, Sun Yat-sen University, Zhuhai 519082, China}
\affiliation{$^2$Sino-French Institute of Nuclear Engineering and Technology, Sun Yat-Sen University, Zhuhai 519082, China}
\affiliation{$^3$Bogoliubov Laboratory of Theoretical Physics, Joint Institute for Nuclear Research, Dubna 141980, Russia}
\affiliation{$^4$Research Institute for Nuclear Problems, Belarusian State University, Minsk 220030, Belarus}

\begin{abstract}
The relativistic vortex states of massive charged particles propagating in non-uniform axisymmetric electric field are studied. Starting from the stationary-state equation after the relativistic Foldy–Wouthuysen (FW) transformation and employing the paraxial approximation, the coupled evolution equations for the beam width, wavefront curvature, and Gouy phase are derived. The equations are solved numerically for a quadratic electrostatic potential, an immersion lens, and an einzel lens. The essential influence of the transverse field on beam evolution is demonstrated. The results provide a relativistic quantum framework for controlling accelerated vortex particle beams using electrostatic fields.
\end{abstract}
\maketitle

\section{Introduction}
Vortex (twisted) states carrying the orbital angular momentum (OAM) are structured wave packets with a quantized projection of the orbital angular momentum on a well-defined propagation axis. The vortex state carries an azimuthal phase factor \(\exp{(il\phi)}\), where \(\hbar l\) (topological charge) is the corresponding eigenvalue of the canonical OAM operator (\(\hat{L}_z\)). It has been shown that such vortex beams or wavepackets can emerge in free electron waves, particularly in electron microscopy. Vortex electrons hold considerable potential for applications across both fundamental and applied physics (see Refs.\cite{3,4,5,6,7,8,9,10,11} and references therein).

However, the limited energy range of vortex electrons and other vortex particles is a key factor restricting their practical applications. Achieving efficient acceleration of these vortex beams would therefore be a very significant step forward. A uniform electric field can basically be used to accelerate a vortex particle beam\cite{12}, whereas axially symmetric electromagnetic lenses can be used to guide and focus an already accelerated beam\cite{13}. Twisted laser beams may also be employed for electron injection and acceleration\cite{14,15}. The generation, acceleration and controlled transport of high-energy vortex particle beams are the important problems of physics of charged vortex particles\cite{16,17,18,19,20,21}. In transmission electron microscopes, twisted electron beams with energies up to \(\sim\) 300 keV and carrying OAM of up to \(\sim\) 1000\(\hbar\) have been experimentally realized\cite{22,23,24,25,26}.

To study twisted electrons accelerated to the relativistic energies, the Dirac equation \cite{15,1,2,Silenko:2018eed,Silenko:2019okz,Zou:2020byc} but not the nonrelativistic Schr\"odinger one should be used. In this work, for twisted states, we apply the normalizable Laguerre-Gauss (LG) wave function to represent the transverse vortex structure, but not the non-normalizable Volkov-Bessel one.

Ref.\cite{2} considered the non-uniform longitudinal electric field and the magnetic field that satisfies the Coulomb gauge. Vortex particles can be accelerated by an electrostatic field. Ref.\cite{1} considered a rigorous treatment of vortex-particle acceleration in a uniform electric field. However, the actual accelerating electric field is usually non-uniform, and the longitudinal non-uniform electric field requires a transverse electric field. The impact of this transverse electric field on the evolution of vortex beam parameters has not yet been studied systematically. The purpose of the present work is to derive the corresponding beam-parameter equations and to quantify these transverse-field effects.

In this paper, we adopt natural units in which \(\hbar=1\) and \(c=1\); however, when presenting numerical estimates, we restore \(\hbar\) and \(c\) explicitly for clarity.

\section{Relativistic quantum mechanics of LG beams in the axisymmetric electric field}
The Dirac Hamiltonian is adopted to describe a particle (without an anomalous magnetic moment) in an electrostatic field\cite{27}:
\begin{equation}\label{eq:-1}
\mathcal{H} = \beta m + \boldsymbol{\alpha} \cdot \boldsymbol{p} + e\mathcal{V},
\end{equation}
where \(m\) is the rest mass of particle, \(e\) is the charge of the particle (\(e=-|e|\)) and \(\mathcal{V}\) is the scalar potential of the electrostatic field, and standard denotations of the Dirac matrices are applied.

Our analysis applies both to the case of an infinite axisymmetric electric field and to the situation where a charged particle beam enters the axisymmetric electric field region from free space. In the latter case, consider that the electric field exists only in the $z>0$ region, and there is no electric field in the $z<0$ region. According to quantum mechanics, the wave function and its first-order derivative are continuous at the boundary.

The relativistic Foldy-Wouthuysen (FW) transformation serves as a convenient formalism for obtaining a transparent Schr\"odinger picture of relativistic quantum mechanics\cite{27,28,29,30,31}. When the weak-field approximation is applied, the relativistic FW transformation for a particle in electrostatic field results in\cite{27}
\begin{equation}\label{eq:0}\begin{aligned}
&H_{FW}  = \beta \epsilon + e\mathcal{V} \\
& + \frac{\mu_0 m}{4} \left\{ \frac{1}{\epsilon(\epsilon + m)} 
, \Bigl[ \boldsymbol{\Sigma} \cdot (\boldsymbol{p} \times \boldsymbol{E} - \boldsymbol{E} \times \boldsymbol{p}) - \hbar \boldsymbol{\nabla} \cdot \boldsymbol{E} \Bigr] \right\}, \\
\end{aligned}\end{equation}
where \(\mu_0=e\hbar/2m\) is the Dirac magnetic moment and \(\epsilon=\sqrt{m^2+\boldsymbol{p}^2}\).

The on‑axis potential \(V(z)\) is assumed to be known for all points on the axis (i.e., at \(r=0\)) and to depend only on $z$. The field at off‑axis point can then be expressed in terms of this on‑axis distribution and its derivatives with respect to $z$, combined with powers of $r$. This near-axis expansion is known as the Scherzer expansion.
\begin{equation} \label{eq:1}\begin{aligned}
\mathcal{V}(z,r) & =\sum_{n=0}^{\infty} \frac{(-1)^n}{(n!)^2}V^{(2 n)}(z)\left(\frac{r}{2}\right)^{2 n} \\
& =V(z)-\frac{1}{4} V^{\prime \prime}(z) r^2+\frac{1}{64} V^{(4)}(z) r^4-\cdots,
\end{aligned}\end{equation}
where $V(z)=\mathcal{V}(z,0)$. Importantly, the potential $V(z)$ can be significantly nonlinear. 

Under the paraxial approximation, $r$ is usually considered to be small. The electric potential may therefore be approximated to the second order in \(r\), i.e., \(\mathcal{V}(z, r)\approx V(z)-\frac{1}{4}V^{\prime \prime}(z) r^2\), the corresponding electric fields $E_z\approx -V^{\prime}(z),E_r \approx \frac{1}{2}V^{\prime \prime}(z) r$. In the general case, electrostatic fields in charge‑free regions obey the Maxwell equations $\boldsymbol{\nabla} \cdot \boldsymbol{E}=0,\boldsymbol{\nabla} \times \boldsymbol{E}=0$. For the case considered here, which is practically relevant, vortex beams satisfy the paraxial condition \(\left(\left|\boldsymbol{p}_{\perp}\right| \ll\left|p_z\right|\right)\), the first-order Eq.~\eqref{eq:0} allows the corresponding second-order equation\cite{27}
\begin{equation}\label{eq:2}\begin{aligned}
&\left[\left( i \frac{\partial}{\partial t} - \mathcal{E} \right)^2 - \boldsymbol{p}^2 - m^2\right] \psi = 0, \\
&\mathcal{E} = e\mathcal{V}(z, r) +\\
&\frac{\mu_0 m}{4} \left\{ \frac{1}{\epsilon(\epsilon + m)} 
, \Bigl[ \boldsymbol{\Sigma} \cdot (\boldsymbol{p} \times \boldsymbol{E} - \boldsymbol{E} \times \boldsymbol{p}) - \hbar \boldsymbol{\nabla} \cdot \boldsymbol{E} \Bigr] \right\}.
\end{aligned}\end{equation}

For stationary solutions (\(H_{FW}\psi_{FW}=\mathbb{E}_0\psi_{FW}\)),
\begin{equation}\label{eq:3}\begin{aligned}
&[(\mathbb{E}_0 - \mathcal{E})^2 - \boldsymbol{p}^2 - m^2]\psi = 0, \\
&\mathbb{E}_0 = \epsilon_0 + eV_0 = \sqrt{m^2 + p_0^2} + eV_0.\end{aligned}\end{equation}
where \(\mathbb{E}_0\) is the conserved total energy. 
The spin-dependent term in the formula for \(\mathcal{E}\) is rather small compared to \(e\mathcal{V}(z, r)\) and can be neglected (\(\mathcal{E} = e\mathcal{V}(z, r)\))\cite{1}.
We define \(V_0=0\), so we obtain
\begin{equation}\label{eq:4}\begin{aligned}
& \boldsymbol{p}^2=\left(\epsilon_0-e\mathcal{V}(z, r)\right)^2-m^2 \\
& =\left(\epsilon_0-eV(z)+\frac{e r^2}{4} V^{\prime \prime}(z)\right)^2-m^2 \\
& =\left[\epsilon_0-eV(z)\right]^2-m^2+2\left(\epsilon_0-eV(z)\right) \frac{e r^2}{4} V^{\prime \prime}(z).
\end{aligned}\end{equation}

In the derivation, only the lowest‑order term in $r$ is retained. The quantity \(P^2(z)=\boldsymbol{p}^2(z,0)\) is introduced to describe the longitudinal acceleration process.
\begin{equation}\label{eq:5}\begin{aligned}
& P^2(z)=\left[\epsilon_0-eV(z)\right]^2-m^2\\
& =p_0^2-2 e \epsilon_0V(z)+e^2V^2(z).\\
\end{aligned}\end{equation}

To proceed to the paraxial equation, we introduce the wave numbers \( K(z) = P(z)/\hbar\) and\( K_0 = P(0)/\hbar \). Here,
\begin{equation}\label{eq:6}
K(z)=K_0 \sqrt{1-\frac{2 e \epsilon_0}{P_0^2} V(z)+\frac{e^2}{P_0^2}V^2(z)}.
\end{equation}

Hence, 
\begin{equation}\label{eq:7}\begin{aligned}
& \boldsymbol{p}^2(z,r)=P^2(z)+\frac{e\left[\epsilon_0-eV(z)\right]V^{\prime \prime}(z)}{2}r^2, \\
&  p(z,r)=P(z) \sqrt{1+\frac{e\left[\epsilon_0-eV(z)\right]V^{\prime \prime}(z)}{2 P^2(z)}r^2} \\
& \approx P(z)\left[1+\frac{e\left[\epsilon_0-e V(z)\right]V^{\prime \prime}(z)}{4 P^2(z)}r^2\right]. \\
\end{aligned}\end{equation}

Therefore, 
\begin{equation}\label{eq:8}
k(z, r) \approx K(z)+\frac{e}{4 \hbar^2} \frac{\left[\epsilon_0-e V(z)\right] V^{\prime \prime}(z)}{K(z)}r^2.
\end{equation}

Defining
\begin{equation}\label{eq:9}
\alpha(z)=\frac{e}{4 \hbar^2} \frac{\left[\epsilon_0-e V(z)\right] V^{\prime \prime}(z)}{K(z)}.
\end{equation}

Then, the derivation of Eq.~\eqref{eq:7} requires that \(K(z)\gg \alpha(z)r^2\) and we obtain
\begin{equation}\label{eq:10}
k(z, r) \approx K(z)+\alpha(z)r^2.
\end{equation}

The equation separates the wave numbers and momentum into a longitudinal acceleration contribution $K(z)$ and a quadratic transverse correction $\alpha(z)r^2$, which originates from the radial electric field. In the paraxial regime, \(\boldsymbol{p}^2 \approx p p_z+\boldsymbol{p}_{\perp}^2 / 2\). Consequently, the equation takes the following form 
\begin{equation}\label{eq:11}
\left(2 \boldsymbol{p}^2+2 i \hbar p \frac{\partial}{\partial z}-\boldsymbol{p}_{\perp}^2\right) \psi=0.
\end{equation}

According to \cite{1}, when \(k\) is merely a function of \(z\) and $r$, the substitution \(\psi=\exp\left[i\int k(z,r)dz\right] \Psi\) into Eq.~\eqref{eq:11} results in
\begin{equation}\label{eq:}
\left[2 \boldsymbol{p}^2+2 i \hbar p \left(ik(z,r)+\frac{\partial}{\partial z}\right)-\boldsymbol{p}_{\perp}^2\right] \Psi=0,
\end{equation}
where $p = \hbar k(z,r)$ and $\boldsymbol{p}_{\perp}^2 = -\hbar^2\nabla_{\perp}^2$, then
\begin{equation}\label{eq:12}
\left[\nabla_{\perp}^2+2 i k(z,r) \frac{\partial}{\partial z}\right] \Psi=0.
\end{equation}

In the axially symmetric electric field under consideration, since the change in momentum in the longitudinal direction is much greater than that in the transverse direction, a substitution similar to that in \cite{1} is adopted.
\begin{equation}\label{eq:16}\begin{aligned}
& \psi=\exp\left[i\int K(z)dz\right] \Psi. \\
\end{aligned}\end{equation}

Substituting into Eq.~\eqref{eq:11} results in
\begin{equation}\label{eq:17}\begin{aligned}
& \left[\nabla_{\perp}^2+2 i k(z,r) \frac{\partial}{\partial z}+2k(z,r)\alpha(z)r^2\right]\Psi =0.
\end{aligned}\end{equation}

The condition \(\frac{K(z)}{\alpha(z)}\gg r^2\) requires \(\frac{4}{|eV^{\prime \prime}(z)|}[\epsilon_0-eV(z)-\frac{m^2}{\epsilon_0-eV(z)}]\gg r^2\). The left side of the inequality has a minimum value of \(\frac{4}{|e|\max|V^{\prime \prime}(z)|}\frac{p_0^2}{\epsilon_0}\). If set, the initial kinetic energy is $20\textrm{ keV}$ and $\max|V^{\prime \prime}(z)|=10 \textrm{ MV}/\textrm{m}^2$, then \(\frac{4}{|e|\max|V^{\prime \prime}(z)|}\frac{p_0^2}{\epsilon_0} \sim 10^{-2}\, \mathrm{m^{2}}\).  In the accelerator, this condition \(\frac{K(z)}{\alpha(z)} \gg r^2\) can usually be satisfied. Then \(2 i k(z,r) \frac{\partial}{\partial z}\approx2 i K(z) \frac{\partial}{\partial z}\) and \(2k(z,r)\alpha(z)r^2 \approx 2K(z)\alpha(z)r^2\). For a kinetic energy of 20 \textrm{keV} and a transverse scale \(r \sim 10^{-7}\,\mathrm{m}\), it is estimated that \(K \sim 10^{11}\,\mathrm{m^{-1}}\) and \(\frac{\partial}{\partial z} \sim \frac{1}{z_R} \sim \frac{2}{K(z)w^2(z)} \sim 10^{3}\,\mathrm{m^{-1}}\). Therefore, \(\alpha(z)r^2\frac{\partial}{\partial z}\) can be ignored compared to \(2K(z)\alpha(z)r^2\). Eq.~\eqref{eq:17} becomes
\begin{equation}\label{eq:18}\begin{aligned}
& \left[\nabla_{\perp}^2+2 i K(z) \frac{\partial}{\partial z}+2K(z)\alpha(z)r^2\right] \Psi=0.
\end{aligned}\end{equation}

The paraxial equation in free space,
\begin{equation}\label{eq:13}
\left[\nabla_{\perp}^2+2 i k \frac{\partial}{\partial z}\right] \Psi=0,
\end{equation}
where \(k\) is a constant, and the equation has the following solution for LG vortex beams:
\begin{equation}\label{eq:14}\begin{aligned}
& \Psi=\mathbb{A} \exp (i \Phi), \quad \int \Psi^{\dagger} \Psi r d r d \phi=1, \\
& \mathbb{A}=\frac{C_{n \ell}}{w(z)}\left(\frac{\sqrt{2} r}{w(z)}\right)^{|\ell|} L_n^{|\ell|}\left(\frac{2 r^2}{w^2(z)}\right) \exp \left(-\frac{r^2}{w^2(z)}\right) \eta, \\
& \Phi=l \phi+\frac{k r^2}{2 R(z)}-\Phi_G(z), \quad C_{n \ell}=\sqrt{\frac{2 n!}{\pi(n+|\ell|)!}},\end{aligned}\end{equation}
where 
\begin{equation}\label{eq:15}\begin{aligned}
w(z) & =w_0 \sqrt{1+\frac{z^2}{z_R^2}}, \quad R(z)=z+\frac{z_R^2}{z}, \quad z_R=\frac{k w_0^2}{2},  \\
&\Phi_G(z) =(2 n+|\ell|+1) \arctan \left(\frac{z}{z_R}\right).
\end{aligned}\end{equation}
Here, $\mathbb{A}$ and $\Phi$ denote the amplitude and phase of the vortex states, respectively. $k$ is the beam wave number, \(w_0\) is the beam waist, \(w(z)\) is the beam width, \(R(z)\) is the radius of curvature of the wave front, \(\Phi_G(z)\) is the Gouy phase, \(z_R\) is the Rayleigh diffraction length, \(L_n^{|\ell|}\) is the generalized Laguerre polynomial and n is the radial quantum number. The spinor \(\eta\) is an eigenfunction of the Pauli operator $\sigma_z$: \(\sigma_z \eta^{ \pm}= \pm \eta^{ \pm}, \eta^{+}=\binom{1}{0},\eta^{-}=\binom{0}{1}\).
\(\Psi\) is not an eigenfunction of the operator \(p_z\). Therefore, the free-space wave function characterizes a beam formed by partial waves with different \(p_z\)\cite{1}.

We now assume that the transverse profile retains the LG functional form while the parameters $w(z), R(z)$, and $\Phi_G(z)$ evolve in response to the external non-uniform accelerating electric field. LG modes remain mutually orthogonal for common values of these beam parameters. Substitute the ansatz into Eq.~\eqref{eq:18}. Combine the terms containing \(\frac{(L_n^{|\ell|})^{\prime}}{L_n^{|\ell|}}\) and those containing \(r^2\) separately, the remaining items are written together (Further detailed derivations can be found in the supplementary material of \cite{1}) and indicate the parameters $(w(z), R(z) ,\Phi_G(z))$ evolution equations
\begin{subequations}\label{eq:19}
\begin{align}
& \frac{k_0}{R(z)} = \frac{K(z)\,w'(z)}{w(z)}, \label{eq:19a}\\
\begin{split}&\left[ \frac{ik_0}{R(z)} - \frac{2}{w(z)^2} \right]^2 
+ iK(z) \left[ \frac{ik_0}{R(z)} - \frac{2}{w(z)^2} \right]' \\  
& +2K(z)\alpha(z) = 0 \end{split},\label{eq:19b} \\
& \Phi'_G(z) = \frac{2(2n + |\ell| + 1)}{K(z)\,w(z)^2}. \label{eq:19c}
\end{align}
\end{subequations}

Eq.~\eqref{eq:19a} is consistent with that of the imaginary part equation of Eq.~\eqref{eq:19b}.
Introducing \( f(z) = \dfrac{ik_0}{R(z)} - \dfrac{2}{w(z)^2} \) leads Eq.~\eqref{eq:19b} to the form
\begin{equation}\label{eq:20}
  f^2(z) + iK(z)f'(z)+ \frac{e\left[\epsilon_0-e V(z)\right]V^{\prime \prime}(z)}{2\hbar^2}  = 0.
\end{equation}
Imposing the boundary conditions \((w(0) = w_0, w'(0) = w'_0)\) and using Eq.~\eqref{eq:19a}, the corresponding expression for $f(0)$ is obtained as \(f(0) = \frac{ik_0 w_0'}{w_0} - \frac{2}{w_0^2}\).
Given the initial conditions (\(w_0,w'_0\)), Eqs.~\eqref{eq:19c} and ~\eqref{eq:20} uniquely determine the subsequent beam evolution. The numerical integration of Eq.~\eqref{eq:20}, starting from the initial value $f(0)$, produces the complex function $f(z)$. Therefore, the beam width and radius of curvature are obtained $w(z)=\sqrt{-\frac{2}{\mathrm{Re} f(z)}},R(z)=\frac{k_0}{\mathrm{Im}f(z)}$.

\section{Beam evolution in an azimuthally symmetric field with a quadratic potential}\label{Penning trap}
An example of an azimuthally symmetric electric field is the Penning trap with a quadratic scalar potential (A complete Penning trap also contains a longitudinal magnetic field):
\begin{equation}\label{eq:30}
\mathcal{V}(z,r)=a(r^2-2z^2), \,\,\, a=\text{const},
\end{equation}

The sign of $a$ determines whether the longitudinal and radial forces are accelerating or decelerating and focusing or defocusing for a particle of a given charge. For the electron, as the charge convention adopted here, the sign is chosen such that a beam propagating along the positive $z$ direction is accelerated and radially focused. This model represents the simplest electrostatic configuration in which longitudinal acceleration is coupled to transverse focusing.

The parameter $a$ is set to \(1.4\textrm{ MV}/\textrm{m}^{2}\). Numerical calculations are then performed to solve the equations derived in Eq.~\eqref{eq:19}.

\begin{figure}[htb]
\centering
\subfloat[\label{fig:0a}]{
\includegraphics[width=0.4\textwidth]{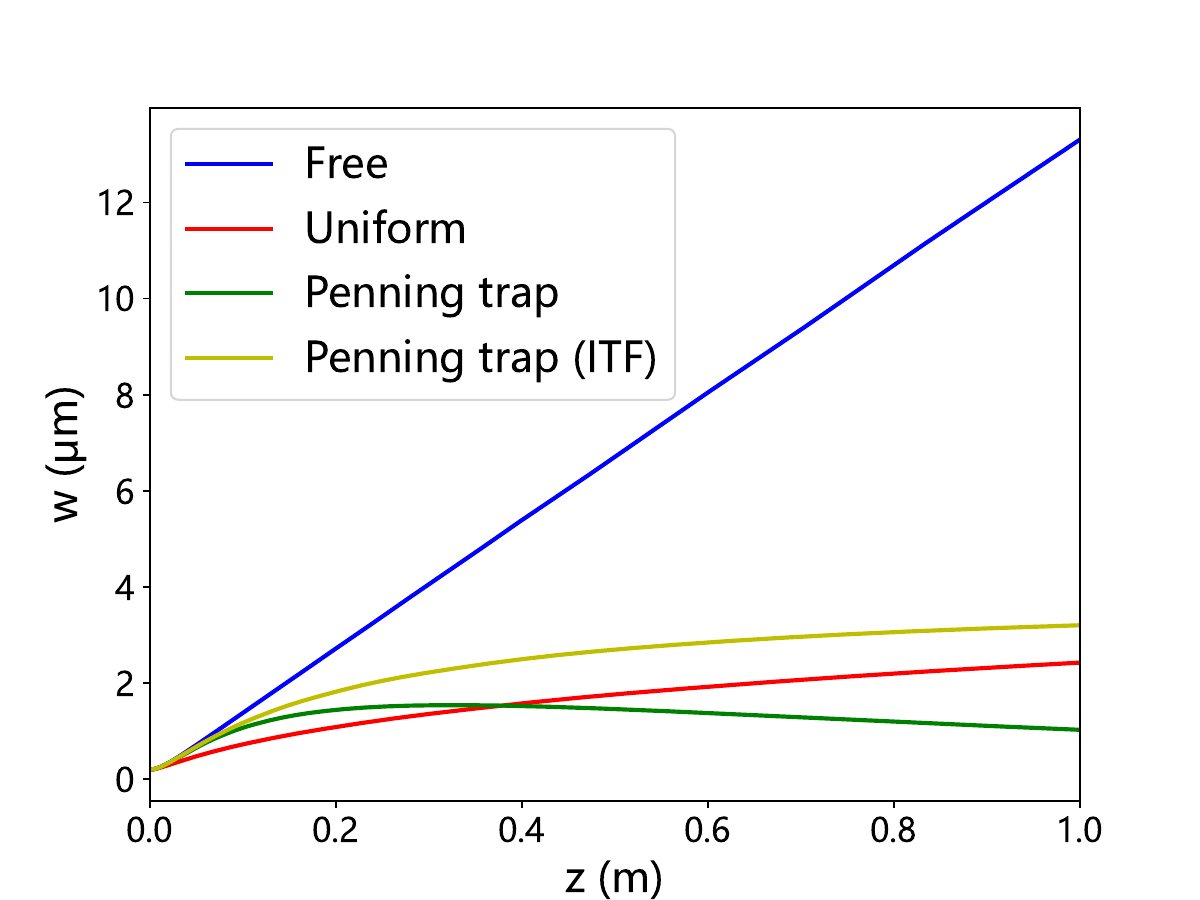}  }
\hfill
\subfloat[\label{fig:0a}]{
\includegraphics[width=0.4\textwidth]{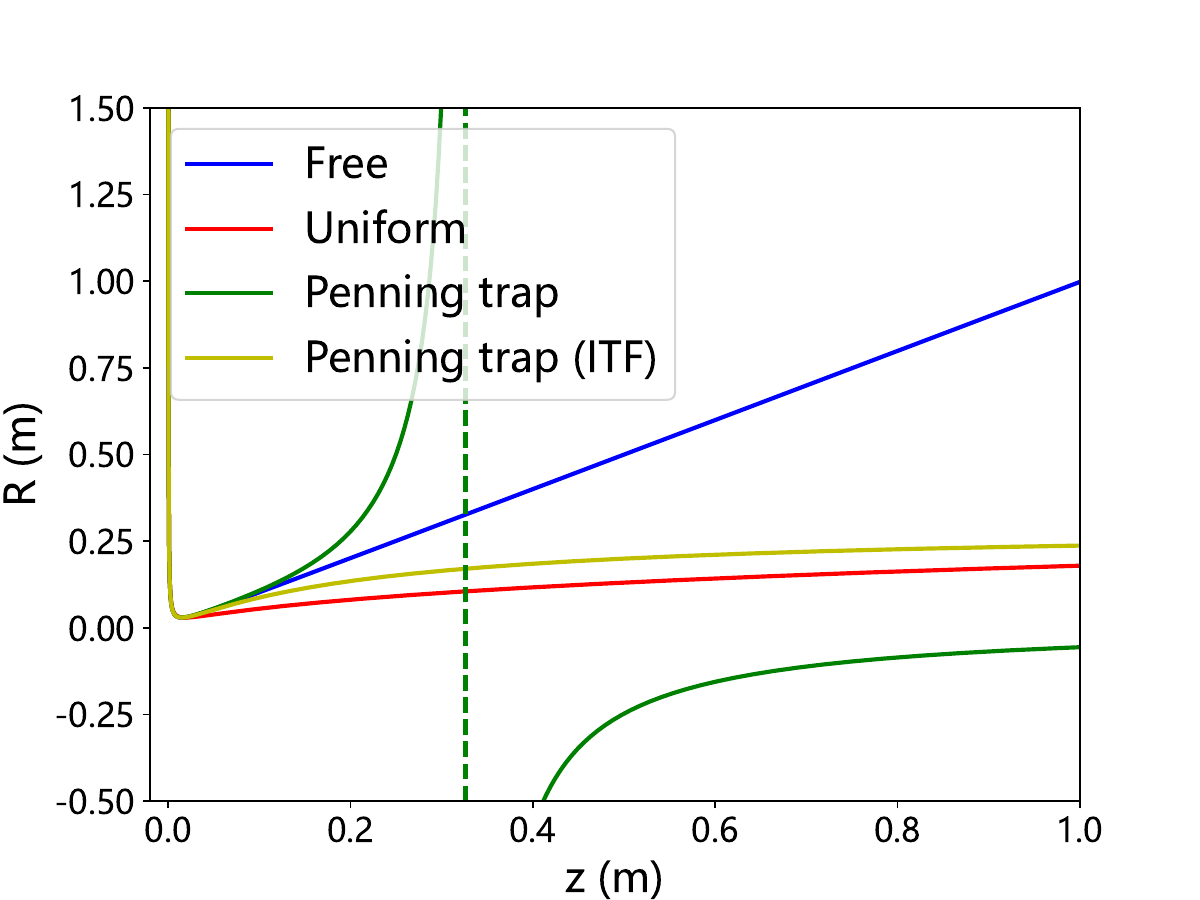}}
 \caption{The initial kinetic energy is $20\textrm{ keV}$, $w=200\textrm{ nm}$, $w'=0$, and different electric fields are considered. (a) Evolution of the beam width $w(z)$. (b) Evolution of the wave front $R(z)$. The blue line corresponds to free space, the red line to the uniform electric field $1.4\textrm{ MV/m}$, the green line to the potential \eqref{eq:30}, and the yellow line to the potential $\mathcal{V}(z,r)=-2az^2$ ignoring transverse field (ITF).}
    \label{fig:0}
\end{figure}

Fig.~\ref{fig:0} compares the beam evolution in free space, in a uniform accelerating field, and in the quadratic electrostatic field, both with and without the transverse component. The difference between the full and longitudinal-only models becomes increasingly pronounced with propagation distance. In the full quadratic field, the radial electric force continuously focuses the beam, causing $w(z)$ to reach a maximum at $z=0.326\textrm{ m}$ and subsequently decrease. At this position, the wavefront radius satisfies $R\rightarrow\infty$, corresponding to a locally flat wavefront. The subsequent change in the sign of $R(z)$ indicates a transition from a diverging wavefront to a converging wavefront. This behavior is qualitatively different from that obtained when only the longitudinal accelerating field is retained.

This example well illustrates how the transverse electric field associated with a non-uniform longitudinal electric field affects the evolution of the beam wave function.

We should note the principal difference between the focusing by electric and magnetic fields. The OAM dynamics is described by the equation
\begin{equation}\label{eq:32}
\frac{d\boldsymbol{L}}{dt} = \boldsymbol{r} \times \frac{d\boldsymbol{p}}{dt} = e \boldsymbol{r} \times (\boldsymbol{E} + \boldsymbol{v} \times \boldsymbol{B}),
\end{equation}
where $\boldsymbol{v}$ is the velocity. As follows from this equation, the radial electric field cannot change the beam OAM. In an axisymmetric electrostatic field, the radial electric force produces no torque about the propagation axis. Moreover, because the vector potential vanishes, the canonical and mechanical OAMs coincide. Therefore, the intrinsic longitudinal OAM of the centered vortex beam, \(L_z=\ell\hbar\), is conserved, although the transverse field can substantially modify the beam width and wavefront curvature. By contrast, a longitudinal magnetic field can modify the mechanical longitudinal OAM during radial expansion or contraction. This circumstance causes the importance to study the behavior of twisted beams in nonuniform electric fields.
\section{Beam parameters evolution in immersion and einzel lens}\label{immersion lens}
\begin{figure*}[htb]
\centering
\subfloat[\label{fig:1a}]{
\includegraphics[width=0.33\textwidth]{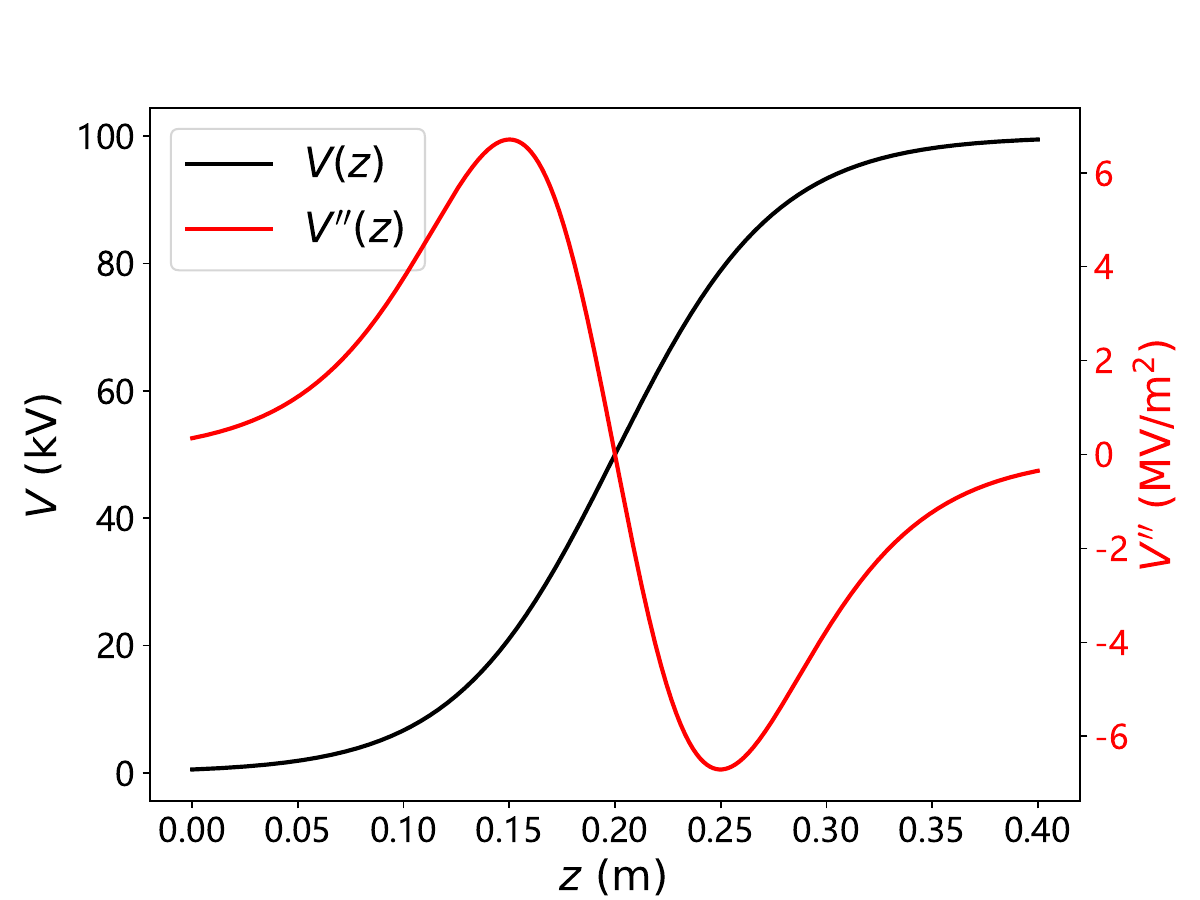}}
\subfloat[\label{fig:1b}]{
\includegraphics[width=0.33\textwidth]{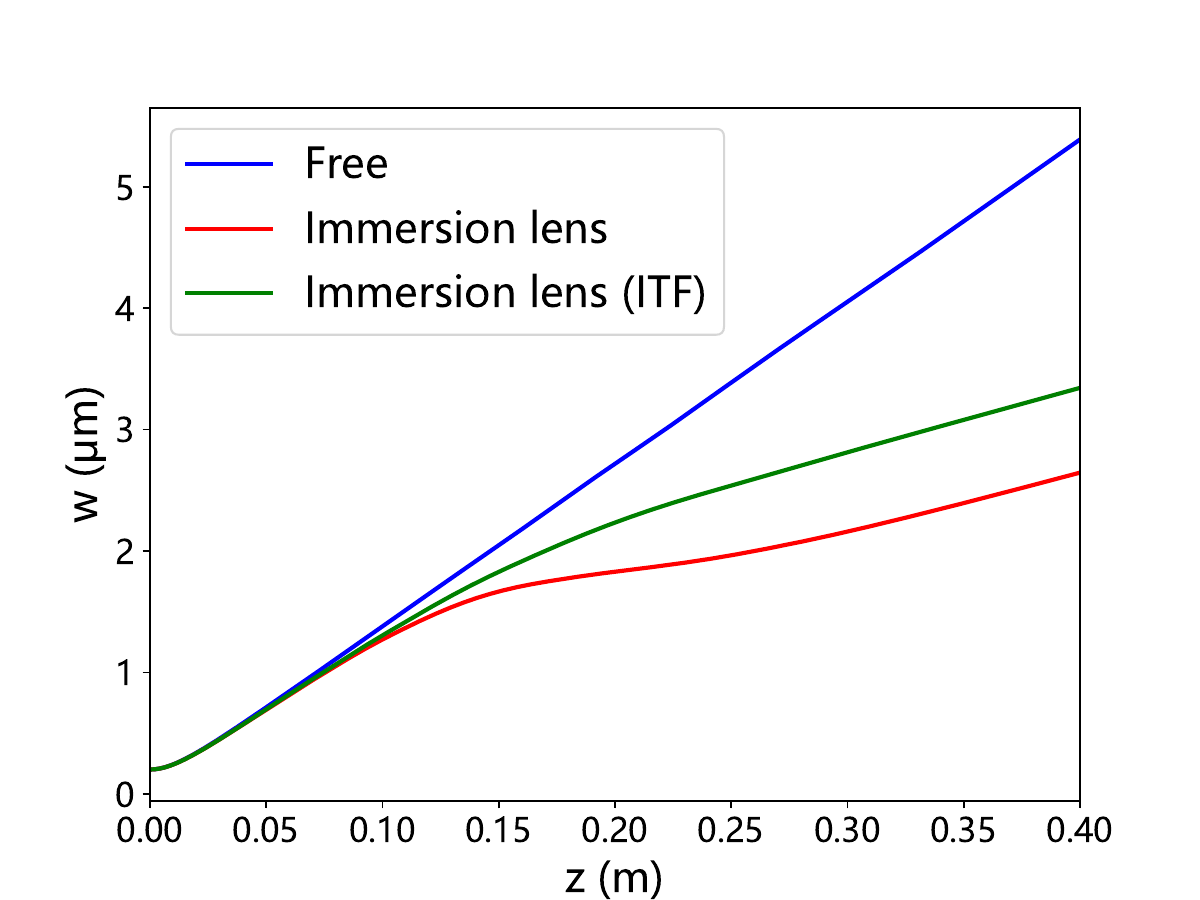}}
\subfloat[\label{fig:1c}]{
\includegraphics[width=0.33\textwidth]{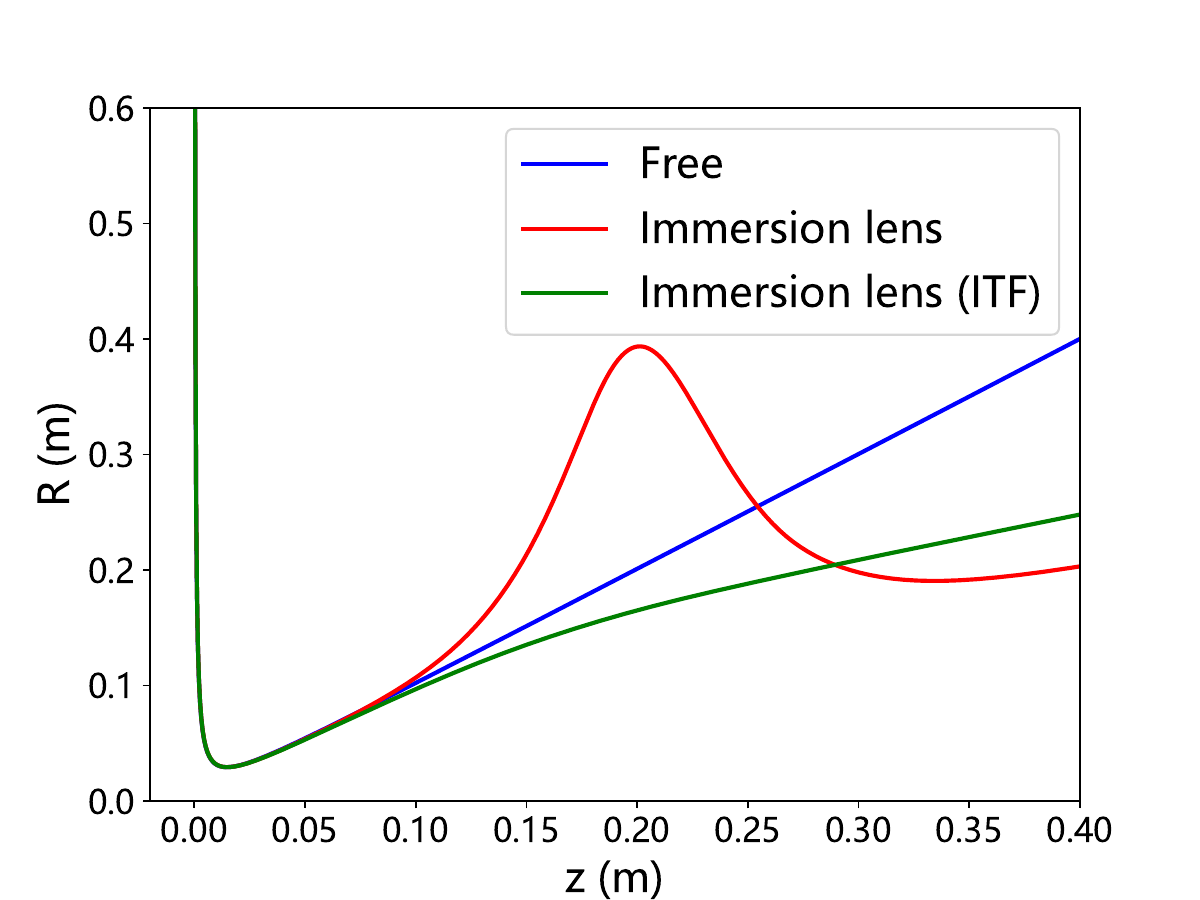}}
\caption{(a) The potential distribution and its second-order derivative function on the axis of the immersion lens model. The initial kinetic energy is $20 \textrm{ keV}$. On the $z=0$ plane, \(w_0=200\textrm{ nm}\), $w'=0$. (b) Evolution of the beam width $w(z)$. (c) Evolution of the wave front $R(z)$. The blue line corresponds to free space, the red line to immersion lens, and the green line to immersion lens that ignores transverse field (ITF).}
    \label{fig:1}
\end{figure*}

\begin{figure*}[htb]
\centering
\centering
\subfloat[\label{fig:4a}]{
\includegraphics[width=0.33\textwidth]{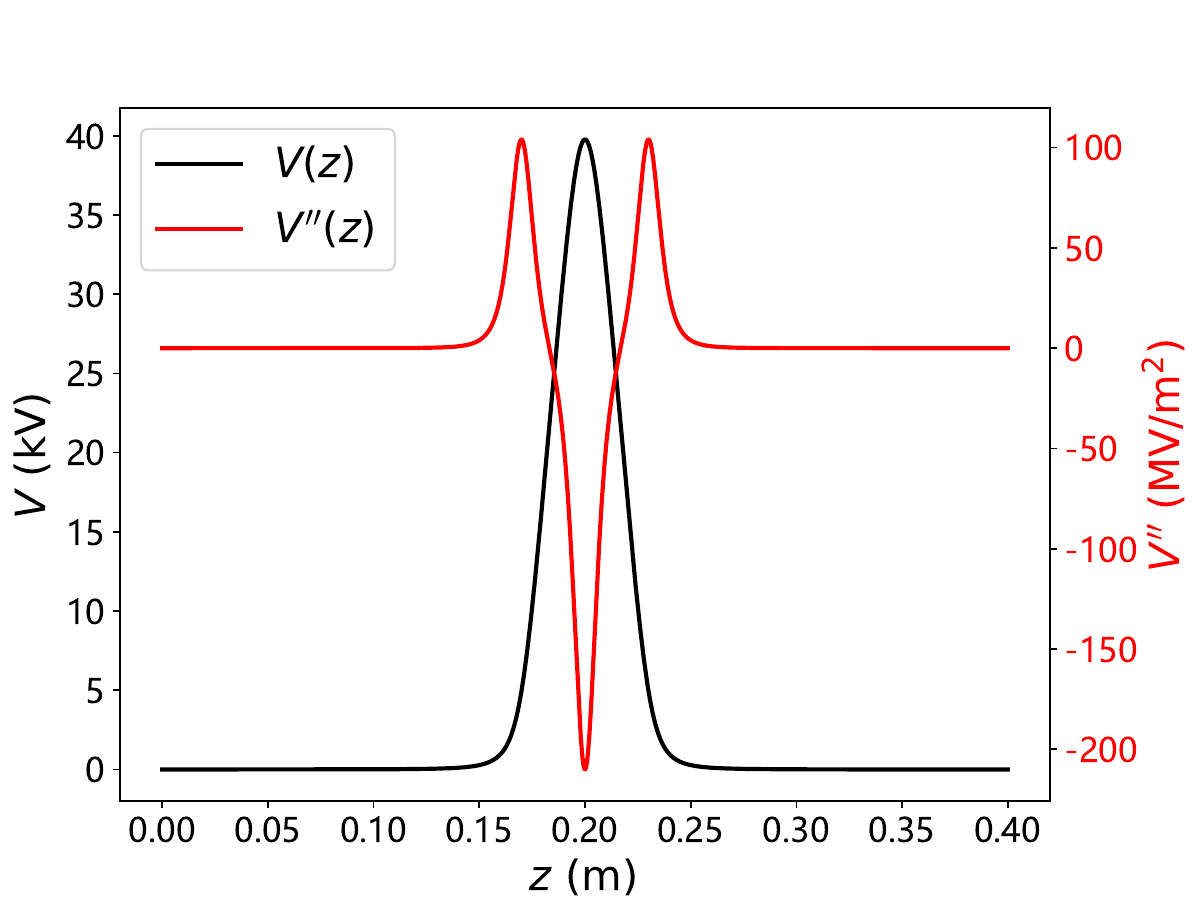}}
\subfloat[\label{fig:4b}]{
\includegraphics[width=0.33\textwidth]{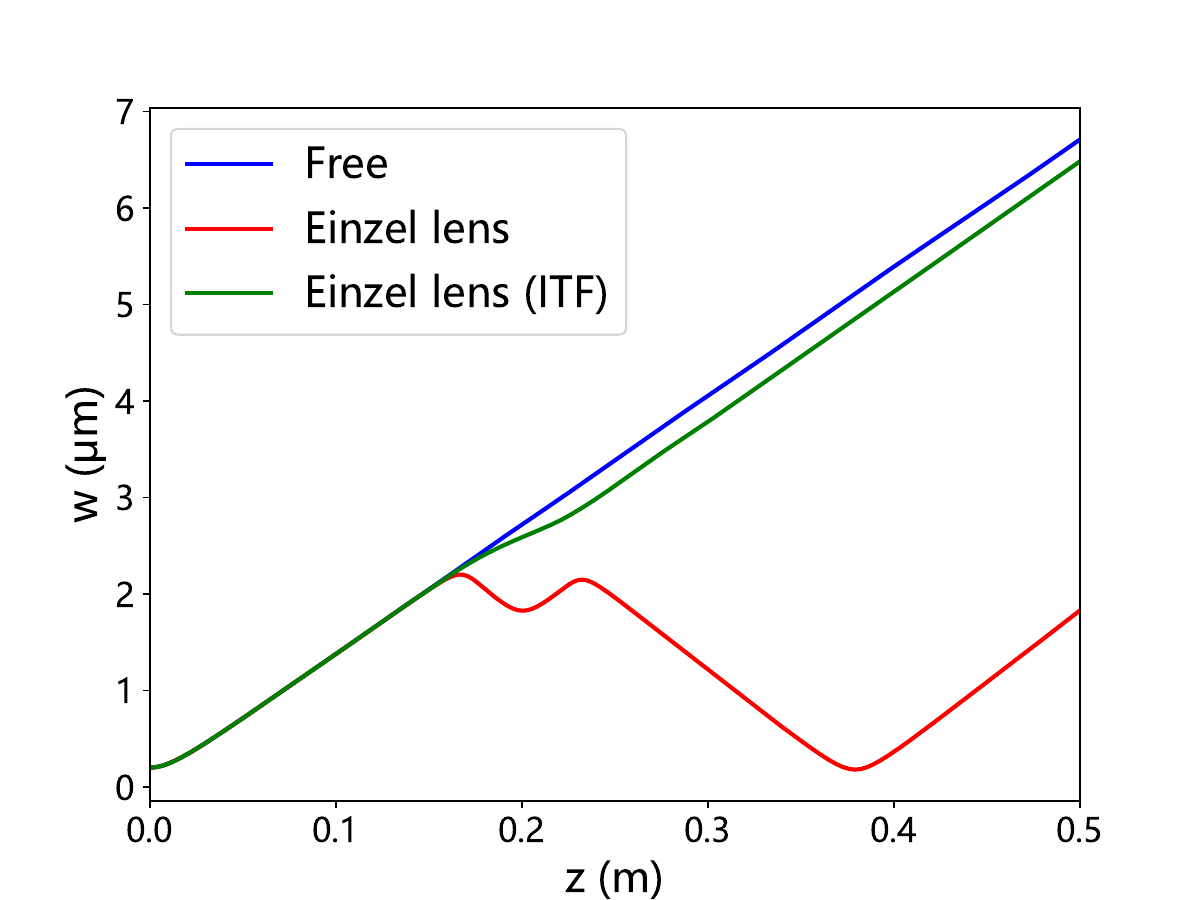}}
\subfloat[\label{fig:4c}]{
\includegraphics[width=0.33\textwidth]{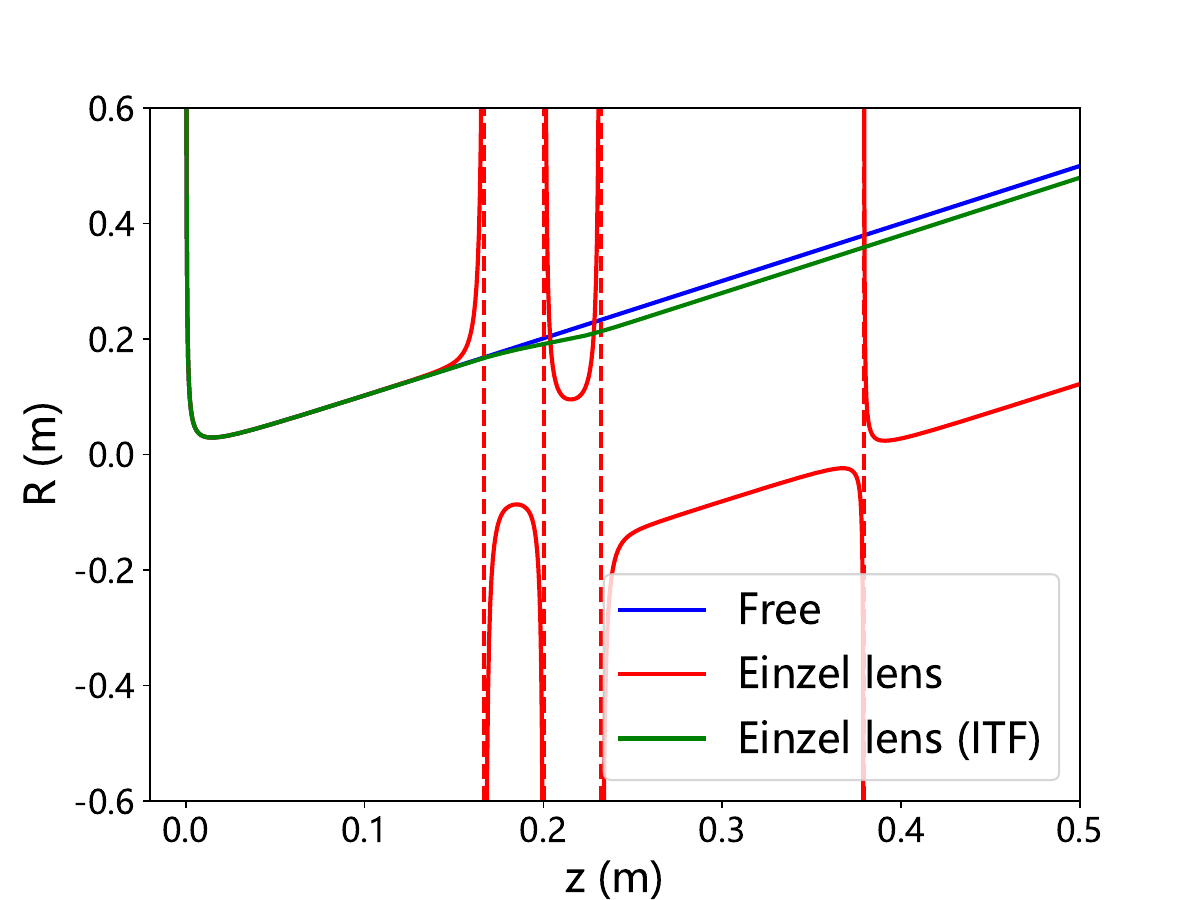}}
\caption{(a) The potential distribution on the axis of the einzel lens model. (b) The evolution of beam width under this potential distribution. (c) Evolution of the wave front $R(z)$. The initial kinetic energy is $20 \textrm{ keV}$. On the $z=0$ plane, \(w_0=200\textrm{ nm}\) and \(w'_0=0\). The blue line corresponds to free space, the red line to einzel lens, and the green line to einzel lens that ignores the transverse electric field (ITF).}
    \label{fig:4}
\end{figure*}
In this section, we consider electrostatic lenses commonly used in beam transmission. A typical example of immersion lenses is the accelerating equal‑diameter bicylinder, in which the two cylindrical electrodes are held at different potentials. The longitudinal variation of the on-axis potential generates a radial electric field and hence a focusing or defocusing force. The on‑axis potential distribution of this lens is
\begin{equation}\label{eq:21}
V(z)= \frac{1}{2}(V_o+V_i)+\frac{1}{2}(V_i-V_o)\tanh \left(\frac{\omega(z-z_m)}{R_m}\right),
\end{equation}
where $V_o$ is the low potential on the left side, $V_i$ is the high potential on the right side, $\omega$ is a parameter of the steepness constant, $R_m$ is the radius of the cylinder and $z_m$ is the coordinate of the geometric center.

For the numerical simulation, we set $V_o=0$, $V_i=0.1\textrm{ MV}$, $\omega=1.32$, $R_m=0.1\textrm{ m}$, $z_m=0.2\textrm{ m}$. The potential on the axis and its second derivative are drawn as shown in Fig.~\ref{fig:1}a.

A numerical method was adopted to solve the Eq.~\eqref{eq:20}, and the curves of the beam width \(w(z)\) as a function of $z$ for different potentials were plotted, as shown in Fig.~\ref{fig:1b}. The curves of the radius of curvature of the wave front \(R(z)\) are presented in Fig.~\ref{fig:1c}.

Fig.~\ref{fig:1b} compares the beam-width evolution obtained with the full field and with the transverse component omitted, revealing a significant influence of the transverse field. Longitudinal acceleration alone suppresses diffraction by increasing the longitudinal momentum \cite{1}. For the chosen charge and potential convention, the accompanying radial force is focusing and further reduces the growth of the beam width.

For the radius of curvature, the transverse electric field has a more significant effect on its evolution (as shown in Fig.~\ref{fig:1c}). When only the longitudinal accelerating field is considered, the radius of wavefront propagating along the z-axis is smaller than that propagating in free space. However, when the transverse field is included, the radius $R(z)$ increases and exhibits a maximum, indicating a marked difference of the wavefront curvature.

The relation between this behavior and the field profile can be understood from Fig.~\ref{fig:1a}. In the first half of the lens, $V^{\prime \prime}$ is positive, and the transverse electric field shows a focusing effect. $R(z)$ increases when focusing, and the wavefront becomes flatter. In the second half, $V^{\prime \prime}$ is negative, and the transverse electric field shows a defocusing effect, with $R(z)$ decreasing.

Using immersion lens model, it is clearly demonstrated that the influence of the transverse electric field in the electrostatic acceleration tube on the evolution of particle parameters cannot be ignored.

The immersion lens features a longitudinal accelerating electric field and a relatively weak transverse electric field. We now consider a primarily focusing device, namely the three‑film einzel lens. The approximate distribution of the electric potential on its axis is
\begin{equation}\label{eq:einzel len}\begin{aligned}
& V(z)= V_o-\frac{V_o-V_c}{\pi L}\Bigg[(z-z_m+L)\arctan\left(\frac{z-z_m+L}{R_o}\right)    \\
&+(z-z_m-L)\arctan\left(\frac{z-z_m-L}{R_o}\right)\\
&-2(z-z_m)\arctan\left(\frac{z-z_m}{R_c}\right)+2(R_o-R_c)\Bigg],
\end{aligned}\end{equation}
where $V_o$ is the potential of the outer electrode, $V_c$ is the potential of the middle electrode, $z_m$ is the coordinate of the center of the lens, $L$ is the axial half-width between two adjacent electrodes, $R_o$ is the radius of the aperture of the outer electrode, and $R_c$ is the radius of the aperture of the middle electrode.

For example, we use $V_o=0$, $V_c=50\textrm{ keV}$, $z_m=0.2\textrm{ m}$, $L=3\textrm{ cm}$, $R_o=1\textrm{ cm}$, $R_c=1\textrm{ cm}$.

The on-axis potential and its second derivative are shown in Fig.~\ref{fig:4a}. Substituting the potential into Eq.~\eqref{eq:20} and solving numerically gives the evolution of the beam width and the wave front (as shown in Fig.~\ref{fig:4b} and Fig.~\ref{fig:4c}).

This model reproduces the focusing action of the einzel lens. If only the longitudinal accelerating electric field is considered, the focusing phenomenon of electrostatic lenses cannot be explained.

From Fig.~\ref{fig:4a} and ~\ref{fig:4b}, it is observed that the transverse field structure of the three‑film einzel lens leads to focusing, defocusing, and subsequent refocusing of the beam width. After passing through the einzel lens, in the region without electric field, Eq.~\eqref{eq:20} still produces a beam width evolution analogous to the free‑space propagation, namely, the width contracts to a beam waist and subsequently expands.

\section{Discussion}
The influence of the transverse electric field on the evolution of beam parameters can be cumulative. Even a very weak transverse electric field, after travelling a sufficiently long propagation distance, will have a significant impact on beam parameters.

The magnitude of the transverse electric field plays a critical role in achieving precise beam focusing and, therefore, impacts the wavefunction parameters of the particle beam, including the beam width. Additional simulations based on Eq.~\eqref{eq:20} show that an excessively strong transverse field, while providing good initial focusing, leads to increased beam divergence as the beam travels further through the lens. Such an enhanced spreading after strong focusing is a common phenomenon in beam optics.

Furthermore, Eq.~\eqref{eq:20} shows that the evolution of beam parameters is the joint effect of beam width, wavefront, on-axis potential, and the second-order derivative of the on-axis potential. The same potential distribution has different focusing effects on beams with different initial beam widths. For more complex cases of the transverse electric field, since $r$ is a small quantity, the electric field can be expanded into the first-order form of $r$ at $r=0$, and the coefficient part related to $z$ in front of $r$ corresponds to the leading-order of Eq.~\eqref{eq:20}. Higher-order terms in the Scherzer expansion would be required to describe spherical aberration and other nonparaxial corrections.

\begin{figure}[ht]
\centering
\subfloat[\label{fig:5a}]{
\includegraphics[width=0.4\textwidth]{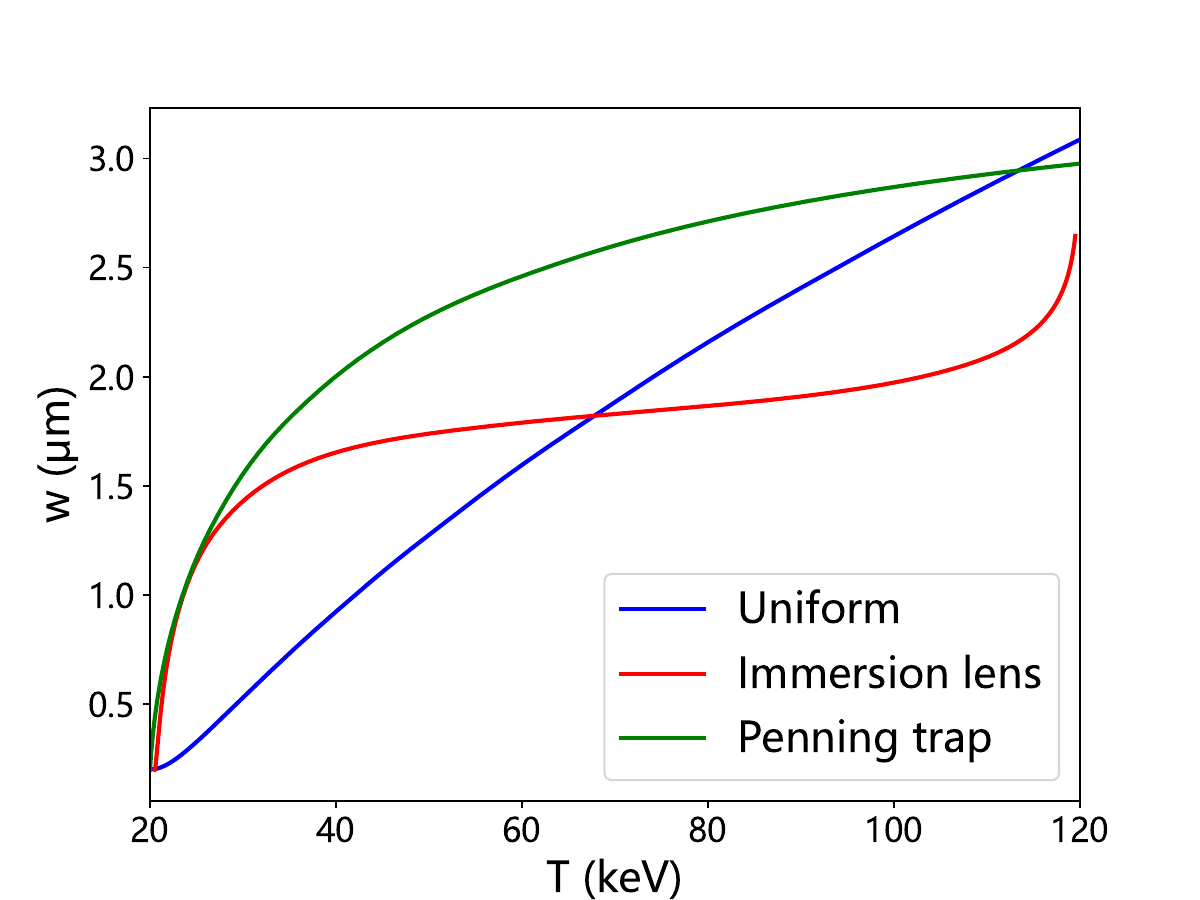} }
\hfill
\subfloat[\label{fig:5b}]{
\includegraphics[width=0.4\textwidth]{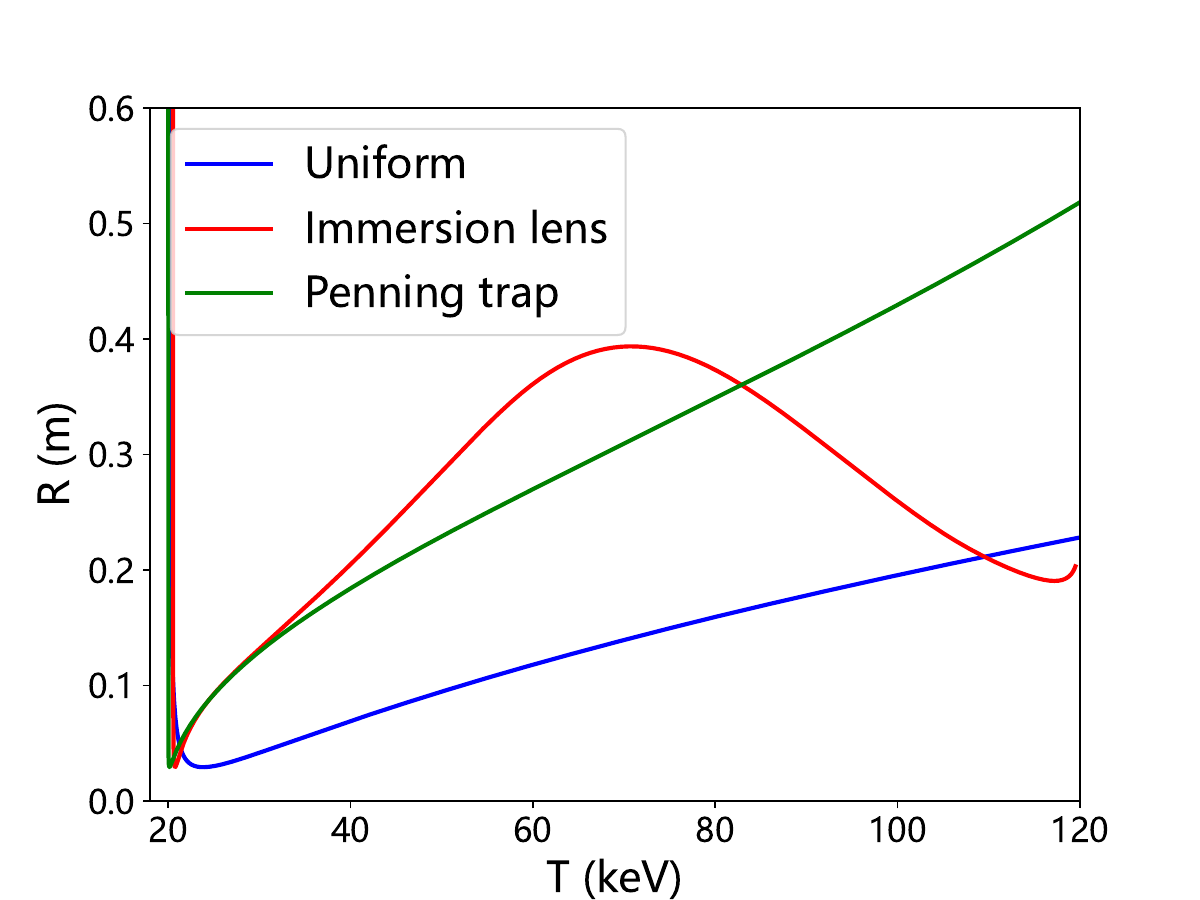}}
\caption{Consider the process of increasing kinetic energy by $100\textrm{ keV}$. (a) The evolution of beam width $w(T)$ as kinetic energy increases. (b) The evolution of wave front $R(T)$ as kinetic energy increases.}
    \label{fig:comparison}
\end{figure}

To better compare the differences among the above accelerating models, a scenario is considered in which the particle kinetic energy increases by $100\textrm{ keV}$ while the immersion lens parameters are kept unchanged. The uniform electric field is set to $E_z=250\textrm{ kV/m}$ and the Penning trap parameter to \(a=312.5\textrm{ kV}/\textrm{m}^{2}\), and the beam is propagated $0.4\textrm{ m}$ along the z-axis. The resulting beam width and wavefront are compared for the same final kinetic energy, as shown in Fig.~\ref{fig:comparison}.

For the selected parameters, the immersion lens model exhibits the least spreading beam width during the intermediate acceleration stage. For the same accelerating energy, a larger second‑order derivative of the potential distribution better suppresses the diffusion of the beam width.

Near the entrance and exit of the immersion lens, the acceleration is weak, so diffraction contributes more strongly to the beam-width growth. However, in the central acceleration region, the rapid increase in longitudinal momentum suppresses diffraction, whereas the radial electric field further modifies the envelope. This behavior demonstrates the unique acceleration property of the immersion lens. Compared with the stable increase of beam width and kinetic energy in a uniform electric field, the increase of beam width in Penning trap becomes slower and slower, while in immersion lens, the beam width diffuses faster in the weak fields on both sides and remains basically unchanged in the large gradient acceleration electric field at the center. These differences illustrate how the spatial structure of the electrostatic field can be used to tailor the propagation of an accelerated vortex beam.

\section{Conclusions}
We have investigated the relativistic paraxial dynamics of charged vortex beams in source-free, axisymmetric, nonuniform electrostatic fields. By combining the Foldy–Wouthuysen representation, the near-axis Scherzer expansion, and an LG-beam ansatz, we derived evolution equations for the beam width, wavefront curvature, and Gouy phase. The transverse electric field associated with the longitudinal potential curvature was shown to produce substantial focusing or defocusing effects that are absent in a longitudinal-only description.

Numerical calculations for a quadratic electrostatic field, an immersion lens, and an einzel lens demonstrate that the radial field can qualitatively change the beam-envelope and wavefront evolution. Within the paraxial approximation, axial symmetry prevents coupling between states with different topological charges, so the vortex quantum number $\hbar \ell$ is preserved during acceleration and focusing. The radial electric field does not change the beam OAM, and vortex structure of the particles is therefore maintained in an axially symmetric electric field. The present formalism can therefore be applied to the analysis and design of axially symmetric electrostatic accelerating and focusing systems for relativistic vortex particle beams.

\section*{Acknowledgments}
The work was supported by the National Key R\&D Program of China (No. 2024YFE0109802), and the National Natural Science Foundation of China (Grants No. 12175320 and No. 12375084).

\bibliography{references}

\end{document}